**Automated detection of business-relevant outliers in e-commerce conversion rate**

Rohan Wickramasuriya [a,*] and Dean Marchiori [b]

[a,*] *SMART Infrastructure Facility, Faculty of Engineering and Information Sciences, University of Wollongong, Northfields Avenue, Wollongong, NSW 2522, Australia*
Email: rohan@uow.edu.au Tel: +61 242 392 546
[b] *Internetrix, 4/85-87 Smith St, Wollongong NSW 2500, Australia*
Email: dean.marchiori@internetrix.com.au

**Abstract**
We evaluate how modern outlier detection methods perform in identifying outliers in e-commerce conversion rate data. Based on the limitations identified, we then present a novel method to detect outliers in e-commerce conversion rate. This unsupervised method is made more business relevant by letting it automatically adjust the sensitivity based on the activity observed on the e-commerce platform. We call this outlier detection method *the fluid IQR*. Using real e-commerce conversion data acquired from a known store, we compare the performance of the existing and the new outlier detection methods. Fluid IQR method outperforms the existing outlier detection methods by a large margin when it comes to business-relevance. Furthermore, the fluids IQR method is the most robust outlier detection method in the presence of clusters of extreme outliers or level shifts. Future research will evaluate how the fluid IQR method perform in diverse e-business settings.

## 1. Introduction

Electronic commerce (e-commerce), which can generally be defined as the process of buying and selling goods or services over the internet [13], has seen an explosive growth worldwide over the last decade. Global Business-to-Consumer (B2C) e-commerce amounted to USD 2.3 trillion in year 2017 [17], while Business-to-Business (B2B) e-commerce is estimated at USD 7.7 trillion by the end of 2017 [16]. Although large businesses have dominated the e-commerce landscape thus far, small and medium enterprises (SMEs) are increasingly benefiting from this rapidly growing economic opportunity [20]. Successful engagement in e-commerce, therefore, has become an essential element for business performance and growth irrespective of the business size and largely irrespective of the business type.

This study revolves around one of the crucial Key Performance Indicators (KPIs) for e-commerce, the conversion rate. Conversion rate in the context of e-commerce is defined as the proportion of the website visitors who make a purchase [8]. Businesses involved in e-commerce strive to achieve and maintain a high conversion rate. Hence, businesses are interested in continuously monitoring the conversion rate on their website, and in particular alerting the management of any abnormality observed in this KPI. If informed in near real time, managers can explore options to bring the conversion rate back to normality in case the abnormality is a negative deviation from the expected conversion rate. On the other hand, timely notification of a positive deviation from the expected conversion rate may be helpful in understanding growth opportunities. For example, a positive anomaly that leads to the discovery that the advertising in a certain outlet draws more leads to the website converting at a higher rate is useful in planning future advertising campaigns.

Detecting marked deviations in the conversion rate can be formulated as a univariate, time series outlier detection problem. Barnett and Lewis [3] define an outlier as an observation or subset of observations which appears to be inconsistent with the other observations in a dataset. In time series data, detecting an outlier is met with an added level of complexity due to the presence of trend and seasonality. For example, there is a possibility of incorrectly identifying a seasonal spike as an outlier in time series data. This added complexity could partly explain the relative sparsity of literature on time series outlier detection. Among existing work, Fox's [9] work is probably the earliest attempt to detect outliers in time series data. His method is based on the general idea that a model representing the structure of time series data could be learned from training data, and this model can then be used to predict the expected next value or values. When the observed value is significantly different from this expected value, the observation could be marked as an outlier. While Fox [9] relied on an auto-regressive model to learn the structure in data, subsequent research has used auto-regressive moving average model [1-2, 21], and auto-regressive integrated moving average model [4, 18] for this purpose. Readers can find a thorough review of literature on outlier detection in Chandola et al. [5].

More recent approaches tend to rely on first decomposing the time series into trend, seasonal and residual components, and then detecting outliers using the decomposed data [10]. A popular decomposition technique has been the Seasonal Trend decomposition procedure based on LOESS or STL for short [7], and more recently its variant [10] known as the 'Twitter method' that uses the median to replace the trend component. If the residual approximates a normal distribution, one can then use Extreme Studentized Deviate (ESD) and its generalized form [14, 15] following a hypothesis testing approach or simply apply the 3 sigma rule to detect outliers. Hochenbaum et al. [10] further develops the outlier detection algorithm by replacing mean and standard deviation in ESD with median and median absolute deviation (MAD) respectively to account for cases with a high percentage of outliers in a dataset. Alternatively, one can rely on Tukey's [19] rule of thumb which is based on the Inter-quartile Range (IQR) to detect outliers as a way to circumvent data distribution assumptions. We discuss these methods in detail in later sections of this paper.

Despite this intriguing existing work, choosing the appropriate combination of methods and their implementation to detect anomalies in conversion rate time series is not straightforward. In this paper, we first describe important characteristics pertinent to conversion time series data, then generate synthetic conversion rate time series that preserve these characteristics, and systematically experiment how various steps of outlier detection techniques perform in practice with conversion data. We then introduce the *fluid IQR,* a technique capable of adjusting the outlier detection sensitivity to improve the business relevance of detected outliers. Finally, we demonstrate the superiority of the fluid IQR using a real dataset obtained from a known e-commerce site.

## 2. Methods

### 2.1 Characteristics of online conversion data

In this paper, we use a slightly different definition for the conversion rate to that used by Fatta et al. [8]. We define the conversion rate as the number of transactions divided by the number of sessions occurring on an e-commerce website in a given period of time. Throughout this paper, we use an hour as that unit time. Below, we summarise key characteristics of e-commerce conversion rate data.

- In the fast-paced e-commerce environment, managers typically look for near real time information. Consequently, the e-commerce conversion rate data that is normally dealt with is of the type sub-daily time series.
- E-commerce conversion data exhibits multiple seasonality. In the case of hourly data, one can expect seasonal periods such as hour of day and hour of week. Usually, the hour of day seasonality is much stronger compared to the day of week seasonality.
- E-commerce conversion rate data usually exhibits trends over medium to long term.
- In addition to the multiple seasonality and trend, conversion rate data carries noise with frequent mild outliers and infrequent strong outliers. Outliers tend to skew the overall distribution of data.
- Conversion rate data are strictly non-negative, although it is common to have zero-valued observations.

From our definition, it is clear that a main driver of the conversion rate is the number of sessions the website has attracted. For instance, low activity on a website close to and after midnight is the reason for near zero conversion rate during this time of day. This is no surprise. However, the other driver of conversions, which is the quality of the leads, is far more important for businesses. For example, low-quality leads drive the number of sessions up, but cause the conversion rate to drop as they make fewer purchases. An unsupervised anomaly detection system should be able to make the distinction between inactive periods and low-quality leads that equally result in low conversion rate. Furthermore, an extra one or two purchases during low activity can dramatically increase the conversion rate, but this is of little value to a business. On the other hand, marginally anomalous conversion rates during high activity periods are critical for businesses as a small change in conversion rate during such periods is associated with a large change in purchases either in the positive or negative direction.

## 2.2 Generating synthetic conversion data

For our experiments, we need hourly time series data for the number of sessions and conversion rate. Generated data should carry certain qualities such as strong intra-day seasonality, mild weekly seasonality, medium term trend, and noise with strong outliers. The general formula we used to generate the conversion rate time series is given in Equation 1.

$$C(\bar{h}) = m_d \left| \sin\left(\frac{h_i \pi}{24}\right) \right| + m_w \left| \sin\left(\frac{h_i \pi}{168}\right) \right| + \frac{h_i}{\bar{h}_{max}} m_t + \varepsilon \tag{1}$$

Where:

$m_d =$ maximum conversion rate in a typical day
$m_w =$ maximum conversion rate in a typical week
$m_t =$ maximum magnitude of any trend in the time series
$\varepsilon$ is an error term $\varepsilon \sim N(0, \sigma^2)$
$h_i$ is the ith hour of the time series
$\bar{h}$ is the hourly vector of the time series

We first generated a synthetic hourly time series of conversion data for a period of three months. Then we randomly introduced about 5% strong outliers that follow a highly skewed distribution. Session numbers were also synthesised following a similar approach. We generated 3 different datasets using the foregoing approach (Table 1). An example synthetic time series generated this way is illustrated in Figure 1.

Table 1. Three synthetic time series datasets used in this study

|  | Trend | Hour of day seasonality | Day of week seasonality |
|---|---|---|---|
| Dataset 1 |  | ✓ |  |
| Dataset 2 | ✓ | ✓ |  |
| Dataset 3 | ✓ | ✓ | ✓ |

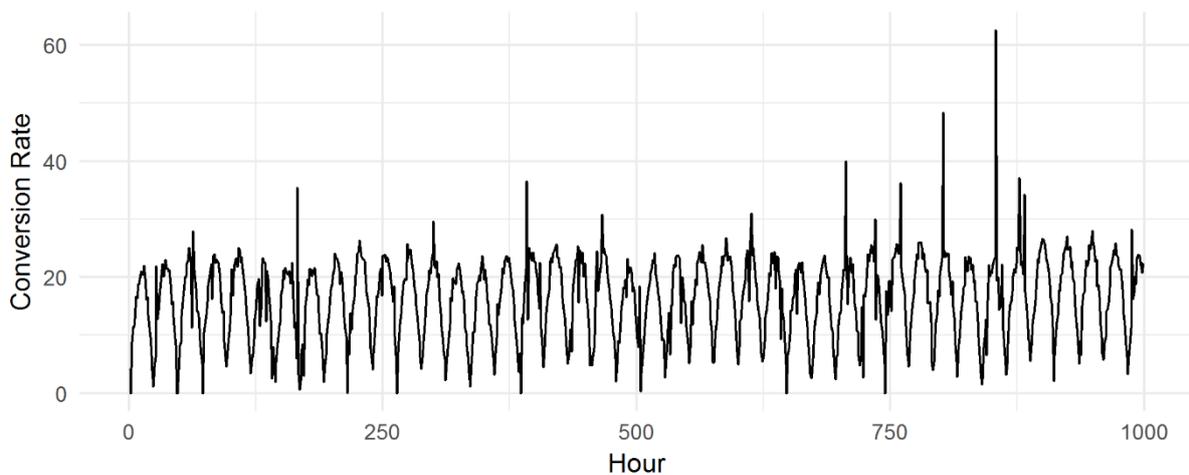

Figure 1. Sample synthetic conversion time series data that carries multiple seasonality, trend and outliers

## 2.3 Systematic experiments

In this study, we chose the more recent approach of first decomposing a time series and then detecting outliers in the residual component. We experimented with two time series decomposition techniques, STL based on LOESS [7] and its recent variant that replaces trend with piece-wise median [10]. We then evaluated the performance of the standard IQR rule [19] as a way of detecting outliers in the residual. Finally, we performed some experiments on adjusting the sensitivity of the IQR rule using session-based weights in order to make the outlier detection more business-relevant. This experiment produced a flexible IQR rule that can transform an unsupervised outlier detection algorithm into a more business relevant one. We call this new rule the *fluid IQR rule*. Following sub-sections discuss these experiments in detail.

### 2.3.1 Time series decomposition

STL based on LOESS, introduced by Cleveland et al. [7], has several advantages over classical time series decomposition methods. Hyndman and Athanasopoulos [11] identify some of these advantages as:

- STL can handle complex, non-linear trends thanks to a LOESS smoother.
- STL's ability to handle any type of seasonality, not just monthly or quarterly seasonality.
- STL can accommodate changing seasonality. Users can even control the rate of change.
- STL is resilient to outliers, i.e. occasional outliers do not affect the estimated trend and seasonal components if robust fitting be used in the LOESS procedure.

At the heart of STL method is a locally weighted regression or LOESS smoother [6] used to derive both trend and seasonal components. Once trend ($T_x$) and seasonal ($S_x$) components are derived for a given time series ($X$), the residual ($R_x$) is calculated as:

$$R_x = X - T_x - S_x \qquad (2)$$

Given that we plan to use the residual component to detect outliers, it is important that the influence of outliers is present in the residual rather than in the trend or seasonal components. STL method has an optional capability to make the trend and seasonal components resilient to outliers. This feature involves defining a robustness weight for each time point that reflects how extreme $R_x$ is. Using the decomposed data, the robustness weight for each time point ($W_t$) is calculated as:

$$W_t = B \left( \frac{|R_{xt}|}{6 \cdot median|R_{xt}|} \right) \qquad (3)$$

Where $B$ is the bisquare function defined as:

$$B(u) = \begin{cases} (1-u^2)^2 & for\ 0 \leq u < 1 \\ 0 & for\ u > 1 \end{cases} \qquad (4)$$

Robustness weight calculation is implemented as an outer loop while the decomposition is implemented as an inner loop in the STL algorithm. In the next iteration of the inner loop, neighbourhood weights are multiplied with the robustness weights. The inner loop and outer loop are repeated for *n* iterations until the convergence is achieved.

We evaluated the STL decomposition with and without the robust LOESS fitting feature, and found that the robust feature in fact improves the decomposition results for our intended purpose of detecting outliers in the residual component. We demonstrate this finding later in the results section. For the decomposition, we used two seasonal periods with frequency 24 and 168, representing hour of day and hour of week respectively. Multiple seasonal decomposition (MSTL) implemented in the R package 'forecast' [12] was used to perform the decomposition. MSTL performs STL in three iterations to extract two seasonal and one trend components.

Furthermore, we evaluated the performance of the STL variant proposed by Hochenbaum et al. [10] for time series decomposition. This variant involves simply replacing the trend component with the median so that the equation 2 becomes:

$$R_x = X - \bar{X} - S_x \qquad (5)$$

Where $\bar{X}$ is the median of the time series.

As we show later in the results section, this STL variant does not perform well in the presence of a substantial trend as is the case with most conversion time series.

*2.3.2 Outlier detection in the residual*

We chose the IQR rule [19] and its variants to detect outliers using the residual component resulting in time series decomposition. Primary reasons for this choice were the simplicity of the method and the fact that the IQR rule does not assume a particular data distribution such as Gaussian. The simplicity associated with IQR means that results can be interpreted easily, thus making the communication of outputs to non-technical audiences less troublesome.

The boxplot method developed by Tukey [19] identifies the 25$^{th}$ percentile as *q1*, the 75$^{th}$ percentile as *q3*, and the difference between *q3* and *q1* as the Inter-quartile range (IQR). He then defines an *inner fence* of the data distribution as the two values *q1 - 1.5 IQR* and *q3 + 1.5 IQR*. He further defines an *outer fence* as the two values *q1 - 3 IQR* and *q3 + 3 IQR*. Observations that fall outside the inner fence are called possible outliers and the observations that fall outside the outer fence are called the probable outliers. That summarises the IQR rule.

During the experimentation, we first applied the standard IQR rule on the residuals to detect potential outliers. The standard IQR rule works well when the data shows no substantial skewness. We transformed the residual component using inverse hyperbolic sine transformation which is defined as:

$$sinh^{-1}(x) = \log(x + \sqrt{x^2 + 1}) \qquad (6)$$

The residual resulting from the decomposition contains negative, positive and in particular zero values. Inverse hyperbolic sine is defined at zero, unlike for example log transformation, making it especially relevant as a method to transform the data at hand.

*2.4 Introducing business relevance to anomaly detection*

E-commerce businesses typically experience high activity and low activity periods on their websites. For example, an online restaurant booking site typically experiences increased traffic during midday and evening, and very low to zero activity after midnight. Additional one or two transactions during typically low activity periods increase the conversion rate dramatically, causing this to be picked up as an anomalous event by a less-careful outlier detection algorithm. Theoretically, this is an anomalous event. However, the business relevance of this particular outlier is negligible. On the other hand, a relatively small change in conversion rate during high activity is of high business relevance, because this is caused by a large change in purchases either in the positive or negative direction. Therefore, it is important that the outlier detection algorithm is more sensitive to changes in conversion rate during high activity periods.

In this paper, we propose a method to adjust the sensitivity of outlier detection based on the intensity of activity observed on an e-commerce website. Recall that in the Tukey's rule the fence is determined by *w.IQR* where *w* = 1.5 yields the inner fence and *w* = 3 yields the outer fence. Our solution lets this factor *w* to vary between $W_{min}$ and $W_{max}$ depending on the level of activity (number of sessions) observed in a given hour. This is achieved by calculating *w* as:

$$w = \frac{S - S_{min}}{S_{max} - S_{min}} \times (W_{max} - W_{min}) + W_{min} \qquad (7)$$

where *S* is hourly sessions, $S_{min}$ is the minimum number of sessions and $S_{max}$ is the maximum number of sessions observed in the hourly time series. In fact, first transforming the session numbers using the inverse hyperbolic sine transformation produces the best results, as the session number can vary between zero and large integers. Once *w* is calculated, the adjusted IQR fence can be calculated simply as:

$$[q_1 - w.IQR \; ; \; q_3 + w.IQR] \qquad (8)$$

We call this new rule *the fluid IQR rule*.

*2.5 Comparing outlier detection methods using real data*

In order to compare the performance of the four outlier detection methods examined in this study, we sourced a real e-commerce dataset from the Google Merchandise Store, which is an e-commerce site that sells Google-branded merchandise. This hourly time series dataset spans over three months, and contains variables such as the number of sessions, transaction and the revenue in dollars. We got the twitter method, STL, MSTL and fluid IQR methods to identify outliers in this conversion rate time series. For each method, we then calculated the total absolute difference in revenue (TADR). In this calculation, hourly median was used as the expected revenue as it is less sensitive to outliers. TADR for all identified outliers *n* by a certain method is given by:

$$TADR = \sum_1^n |r - \bar{r}| \qquad (9)$$

where *r* is the revenue associated with that outlying observation and $\bar{r}$ is the median revenue for that particular hour of week.

## 3. Results and Discussion

*3.1 Systematic experiments*

Note our approach of first decomposing the time series data and then detecting outliers using the remainder component. For the experiments, we first used the time series dataset with only a single seasonality. Then we moved onto the dataset with trend and single seasonality. Finally, we evaluated the performance of decomposition techniques using the time series dataset with trend and multiple seasonality. We started the experiments by evaluating three variants of a decomposition algorithm: STL with no robust LOESS fitting feature, STL with robust LOESS fitting feature and the Twitter method where piece-wise median replaces trend. Figures 2, 3 and 4 illustrate the outputs produced by these three variants applied on the dataset with single seasonality only.

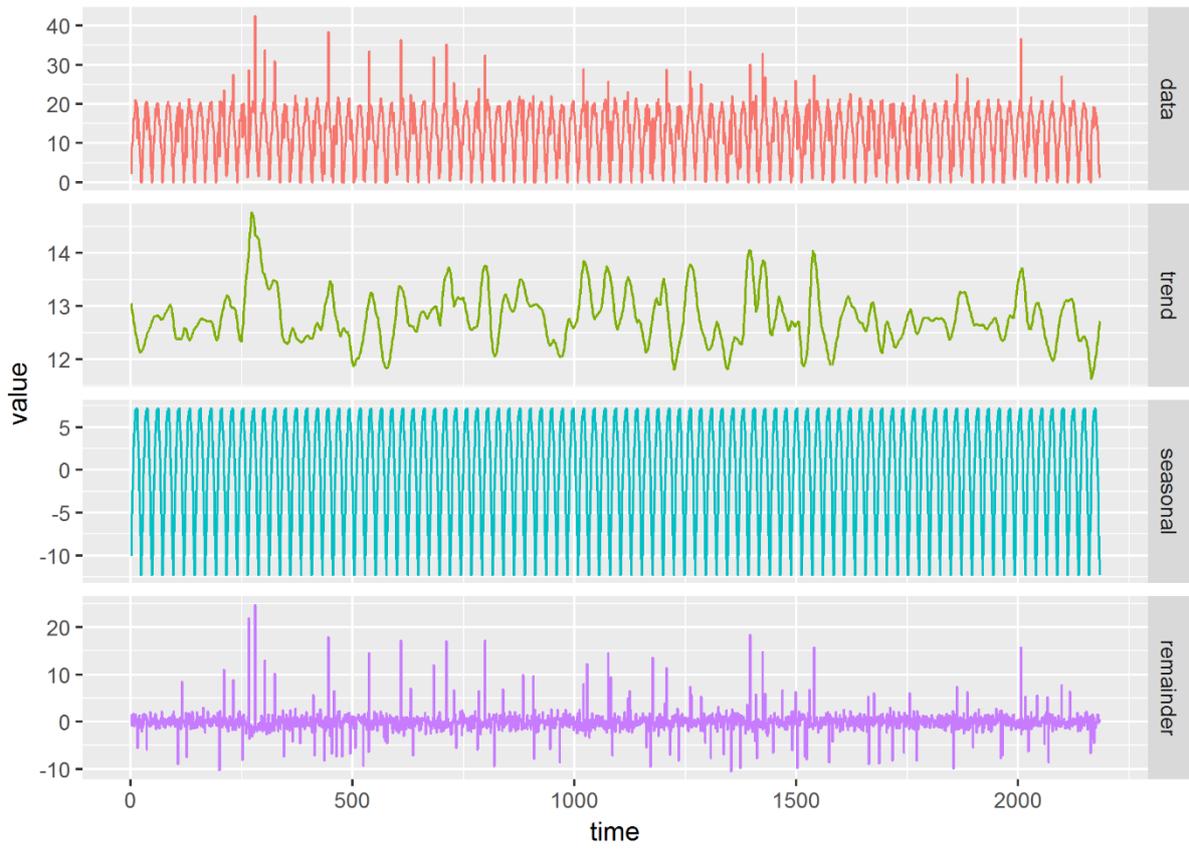

Figure 2. Components of the decomposed time series without robust LOESS fitting

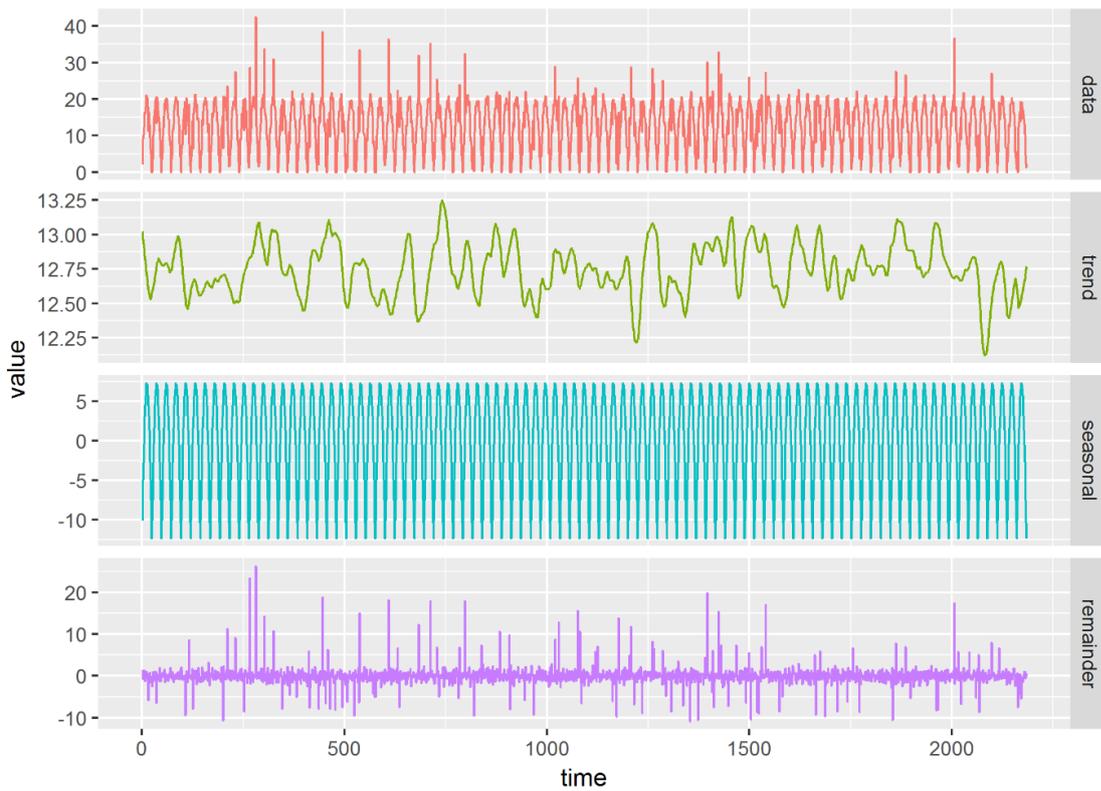

Figure 3. Components of the decomposed time series with robust LOESS fitting

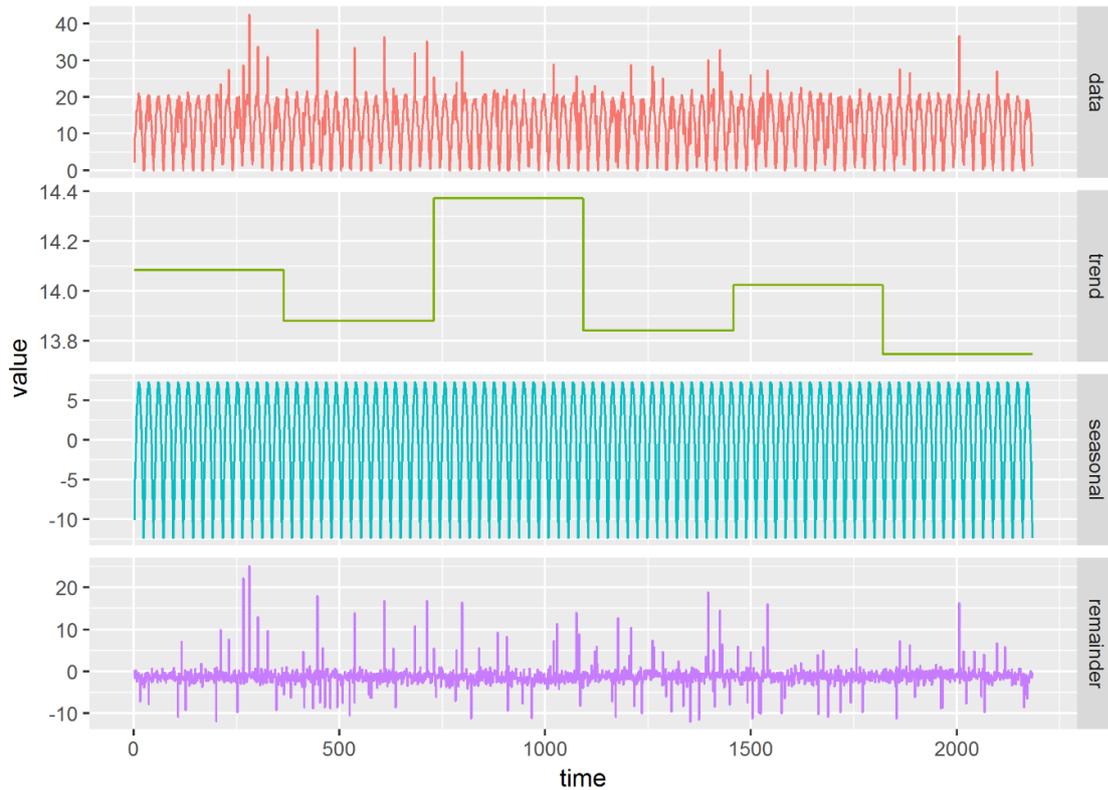

Figure 4. Components of the decomposed time series produced by replacing the trend with a piecewise median

Careful comparison of Figures 2 and 3 reveal that the robust feature has managed to push most of the outlyingness into the remainder component. This is evident from the smaller range associated with the trend component and the larger range associated with the remainder component in the output of STL with robust LOESS fitting compared to those in the outputs generated by STL without the robust feature. Also, note that the robust feature has ensured that the trend does not follow the same spikes as the strong outliers observed in the series. Nonetheless, robust LOESS fitting still leaks some noise into the trend component. On the other hand, visual inspection suggests that the Twitter method that uses piece-wise median as trend (bi-weekly in this case) has produced a comparable remainder component as the robust STL. To empirically evaluate the performance of these three variants, we then used the remainder components to detect outliers using the standard IQR rule. The standard IQR rule uses the outer fence to isolate the outlying observations. Note we had about 120 randomly inserted, labelled outlying points that were used in this evaluation as groundtruth. Figure 5 illustrates the different levels of success the IQR rule has registered when three remainder components are used as inputs. Table 2 compares the same performance using classical indices, accuracy, sensitivity and specificity.

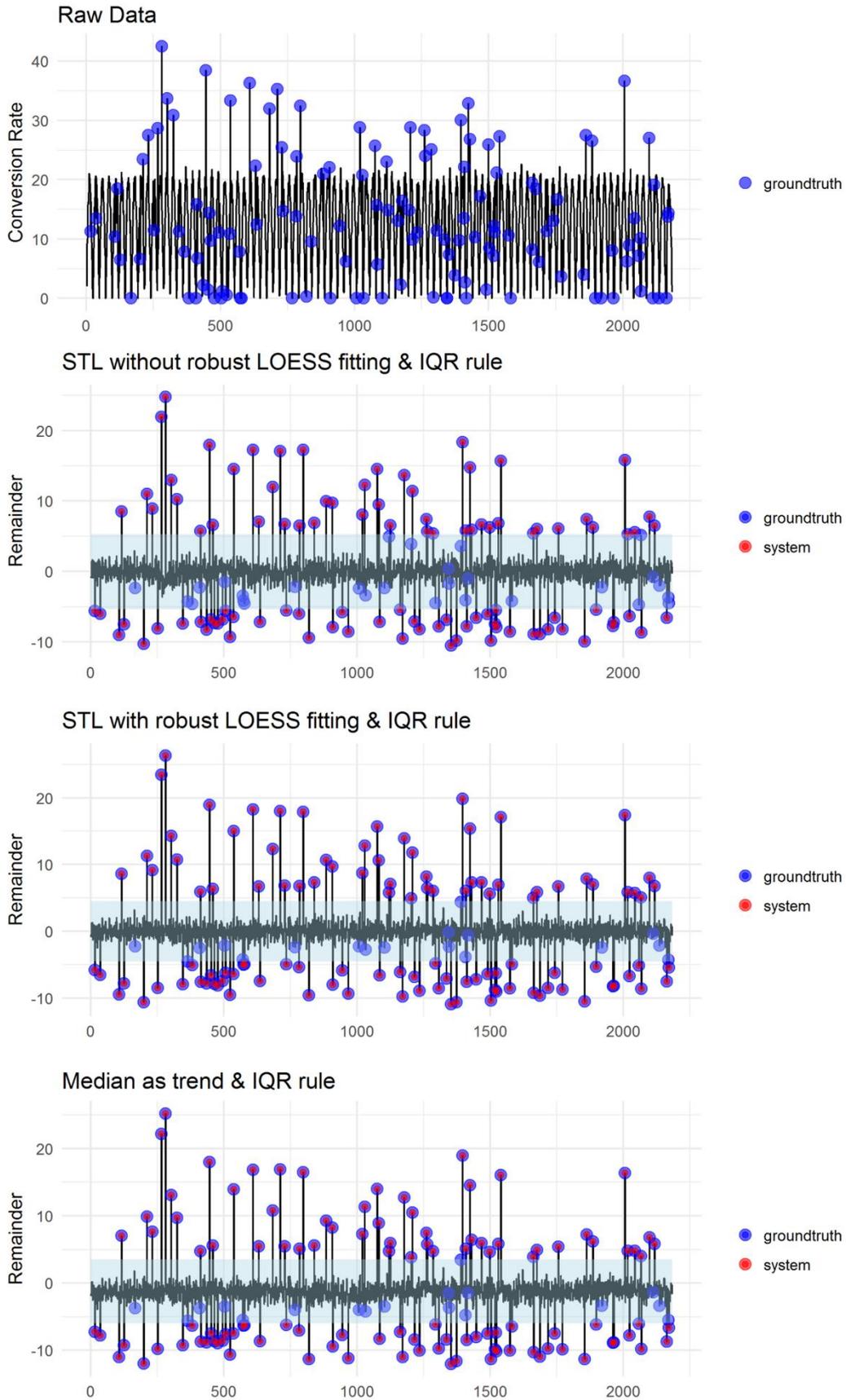

Figure 5. Time series decomposition as a precursor to outlier detection: performance of three decomposition variants

Table 2. Time series decomposition techniques and associated outlier detection accuracy

| Decomposition Method | Accuracy | Sensitivity | Specificity |
|---|---|---|---|
| STL (robust=false) | 0.9872 | 0.7823 | 1 |
| STL (robust=true) | 0.9918 | 0.8605 | 1 |
| Twitter | 0.9918 | 0.8605 | 1 |

The positive case in the evaluation shown in Table 2 and subsequent evaluations is an observation being an outlier (outlier = yes). Sensitivity measures the proportion of actual outliers the system has correctly identified as outliers, while specificity measures the proportion of actual non-outliers (normal observations) the system has correctly identified as non-outliers. Our results suggest that STL decomposition with robust feature and Twitter method perform equally well as a precursor for outlier detection in time series data with single seasonality and no trend. STL without robust feature exhibits significantly inferior performance. With this knowledge, we discarded the STL without robust LOESS fitting as a decomposition technique for conversion time series data. We then performed the same experiment with the second time series dataset that carries a mild trend and single seasonality, using the two remaining decomposition techniques. Figures 6 and 7 depict the outputs of the two decomposition techniques.

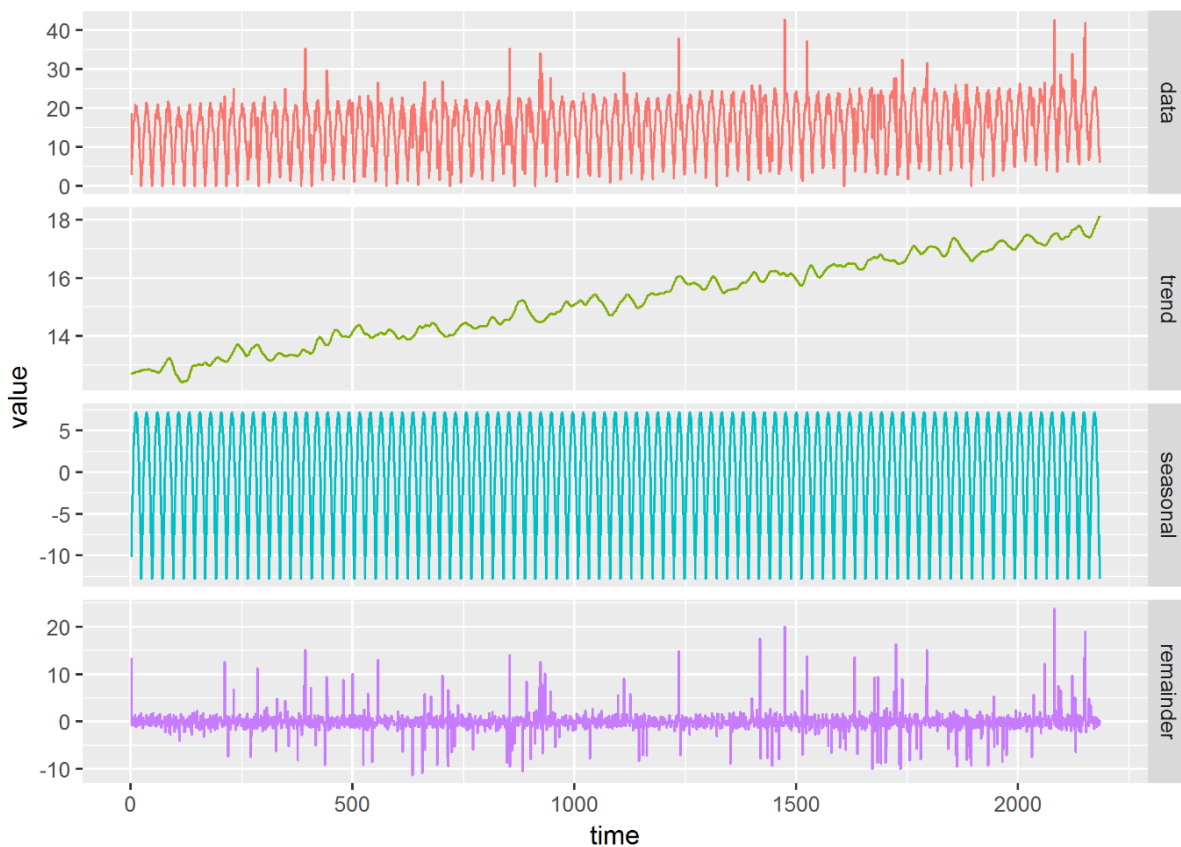

Figure 6. Decomposition of conversion time series data containing trend and single seasonality using STL with robust LOESS fitting

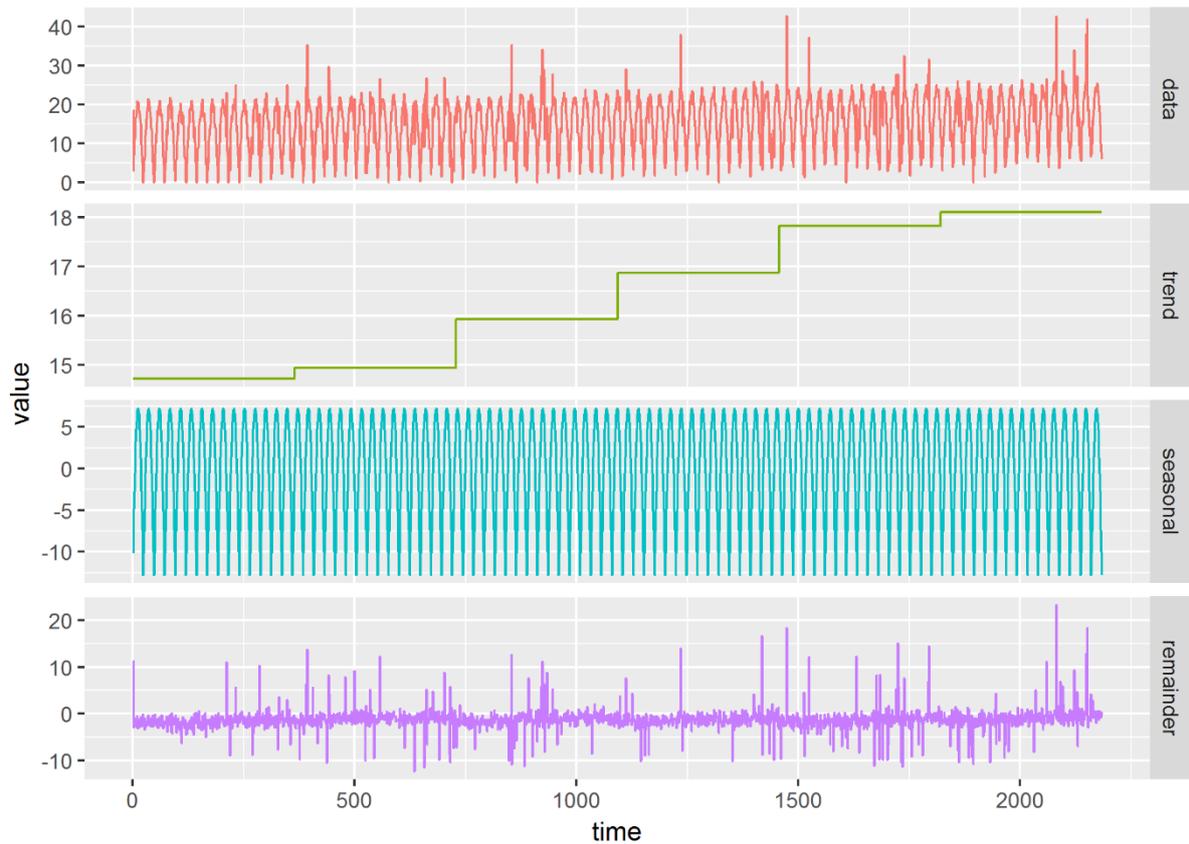

Figure 7. Decomposition of conversion time series data containing trend and single seasonality using Twitter method (median span = 2 weeks)

STL with robust LOESS fitting has clearly isolated the trend and hour of day seasonality. As a result, the remainder seems to follow no pattern. Twitter method also is capable of clearly isolating the hour of day seasonality, but as a result of the piece-wise median replacing trend, the remainder has ended up carrying some trend information. This is undesirable given our purpose of detecting outliers using the remainder component. Figure 8 visually compares and contrasts the outliers detected following the application of two decomposition techniques. Again, we applied the standard IQR rule on the remainder to produce this output. Table 3 provides the accuracy measures for the two methods.

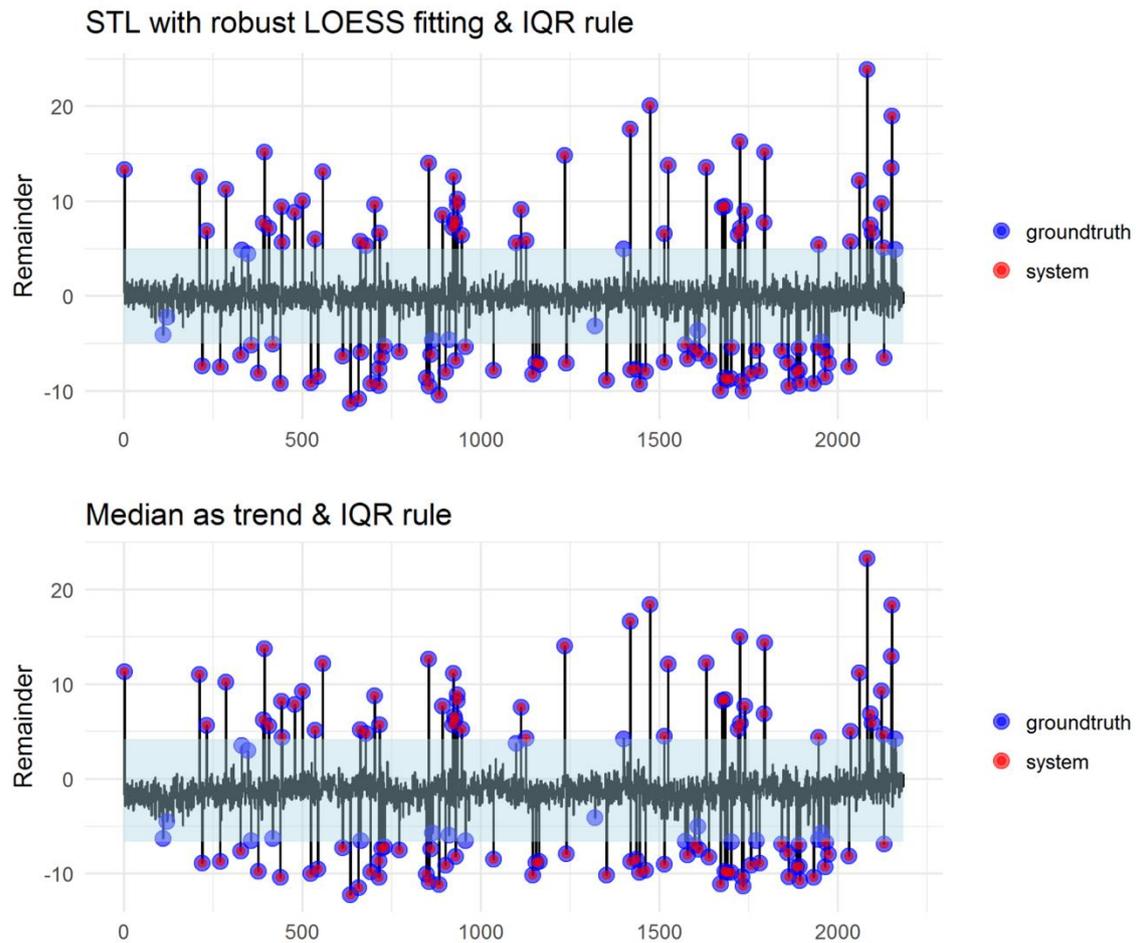

Figure 8. Outlier detection performance of two methods: STL with robust LOESS fitting and Twitter method

Table 3. Single seasonality + trend

| Decomposition Method | Accuracy | Sensitivity | Specificity |
|---|---:|---:|---:|
| STL (robust=true) | 0.995 | 0.9154 | 1.00 |
| Twitter | 0.9908 | 0.8462 | 1.00 |

It is evident from Table 2 that STL with robust feature clearly outperforms the Twitter method, particularly in terms of sensitivity, which is the key performance measure in this evaluation.

Conversion time series data usually contains mild trend over the medium term, as well as multiple seasonality. Using the third dataset, we then evaluated the suitability of STL with robust LOESS fitting and MSTL with robust LOESS fitting [11] as precursors to outlier detection in such data. Figures 9 and 10 illustrate the decomposition outputs of the said two algorithms.

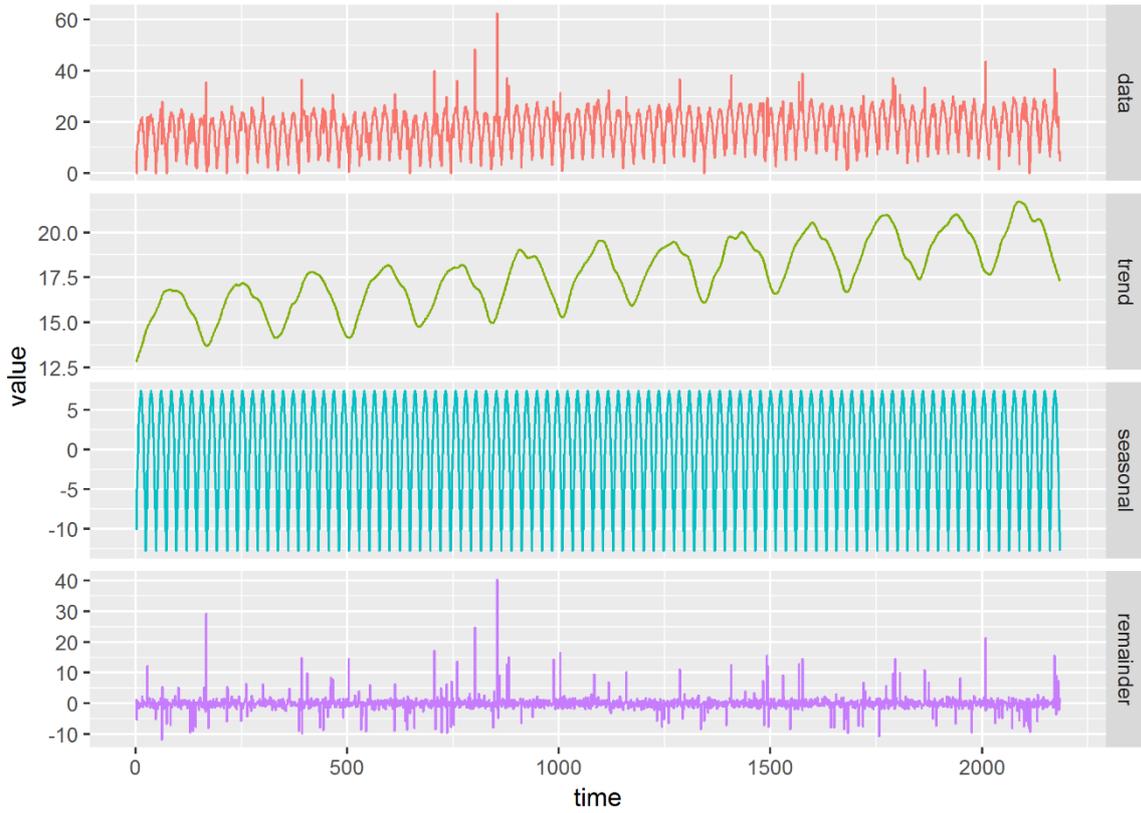

Figure 9. Decomposition of conversion time series data containing trend and multiple seasonality using STL with robust LOESS fitting

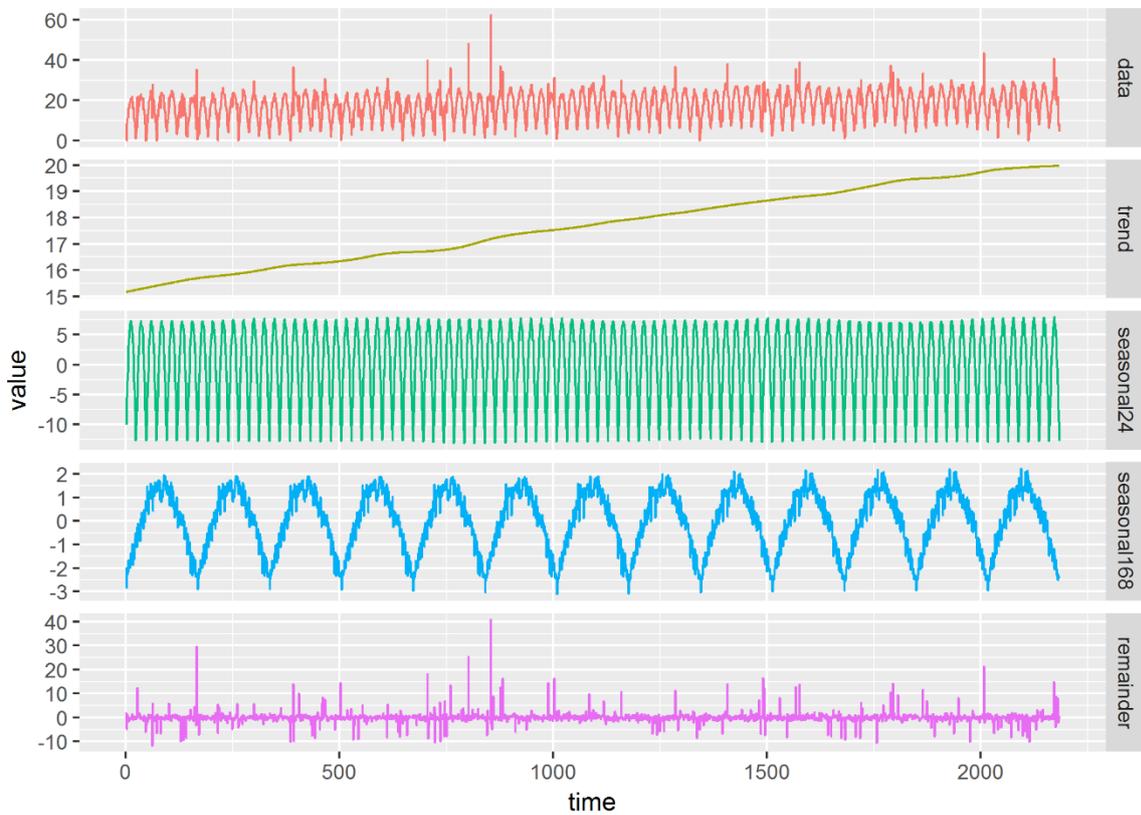

Figure 10. Decomposition of conversion time series data containing trend and multiple seasonality using MSTL with robust LOESS fitting

Interestingly, STL with robust LOESS fitting method has captured the second seasonality in the trend component, while MSTL has produced a much cleaner linear trend, intra-day seasonality and hour of week seasonality. From an outlier detection point of view, all what we care about is a remainder component that carries all the outlyingness. Hence, one could argue that the STL with robust LOESS fitting is equally suitable for this task. However, Figure 11 and Table 4 provide enough evidence to declare that MSTL with robust LOESS fitting is the best decomposition technique for sub-daily conversion time series data with trend and multiple seasonality, and for the intended purpose of detecting outliers.

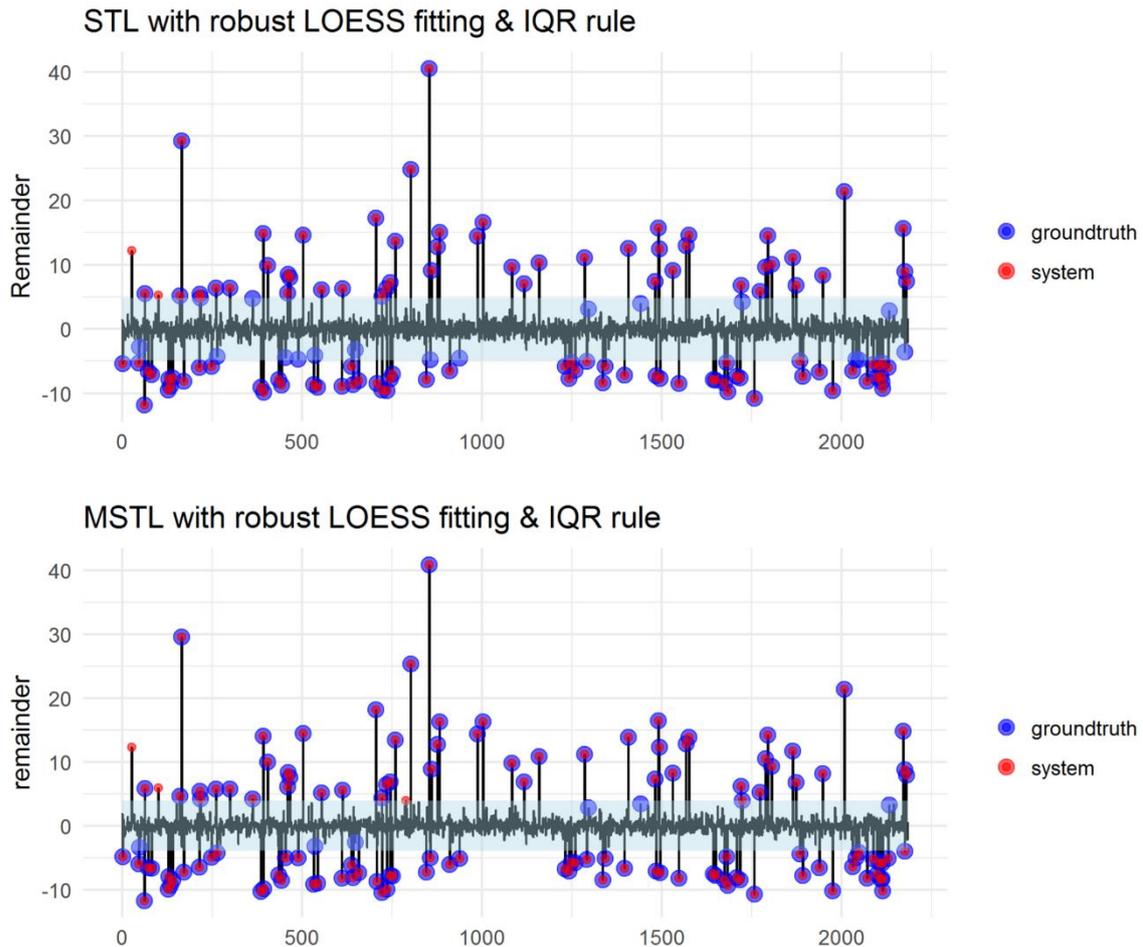

Figure 11. Outlier detection performance of two methods: STL with robust LOESS fitting and MSTL with robust LOESS fitting

Table 4. STL vs MSTL

| Decomposition Method | Accuracy | Sensitivity | Specificity |
|---|---:|---:|---:|
| STL | 0.9918 | 0.875 | 0.999 |
| MSTL | 0.9959 | 0.953 | 0.999 |

Having chosen MSTL with robust LOESS fitting as the preferred decomposition technique, we then explored how we could transform the time series decomposition and IQR rule to be more business relevant, while retaining the desired unsupervised characteristic. Recall that our objective here is to increase the sensitivity of the outlier detection during active periods and to decrease the sensitivity during inactive periods of day. We can achieve this by letting the fence in the IQR rule to dynamically change in response to the level of activity observed on the website. To arrive at this end goal, we first transformed the remainder using inverse hyperbolic sine transformation (Figure 12).

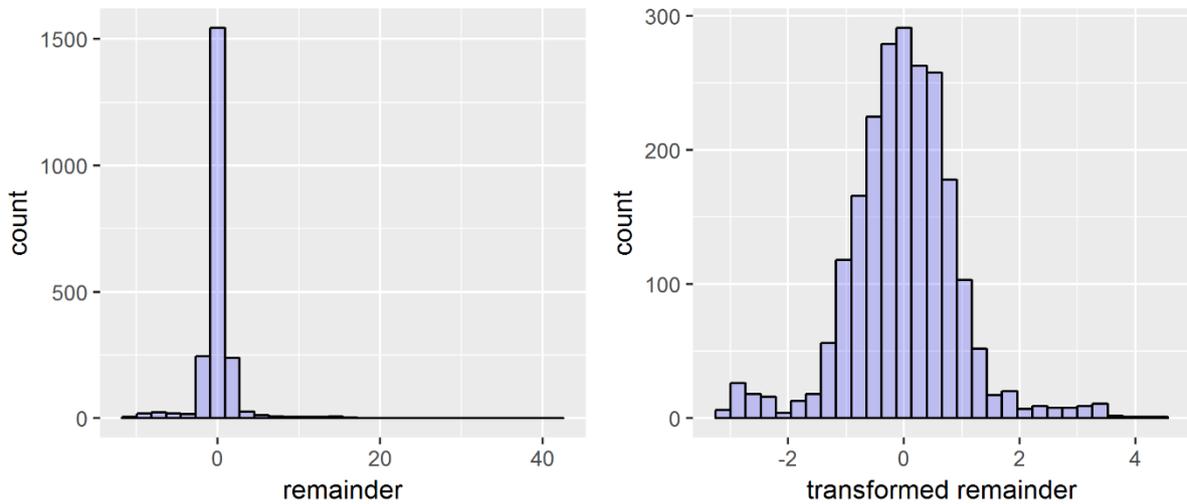

Figure 12. Transforming skewed remainder into approximate normal using inverse hyperbolic sine transformation

We then applied the IQR rule, but with the inner fence (1.5 IQR), to the transformed remainder. This produced nearly the same level of outlier detection performance, in terms of accuracy (0.9954), sensitivity (0.938) and specificity (0.999), as the outer fence applied to the raw remainder. From here, we let the fence to change between 3 IQR and 1.5 IQR depending on the number of sessions, with the highest value of 3 IQR coinciding with the lowest number of sessions and 1.5 IQR coinciding with the highest number of sessions, during time period considered (e.g. 3 months). Figure 13 shows a subset of the raw data and transformed remainder series with both hand labelled and system detected outliers using this new IQR rule we choose to call *the fluid IQR rule*.

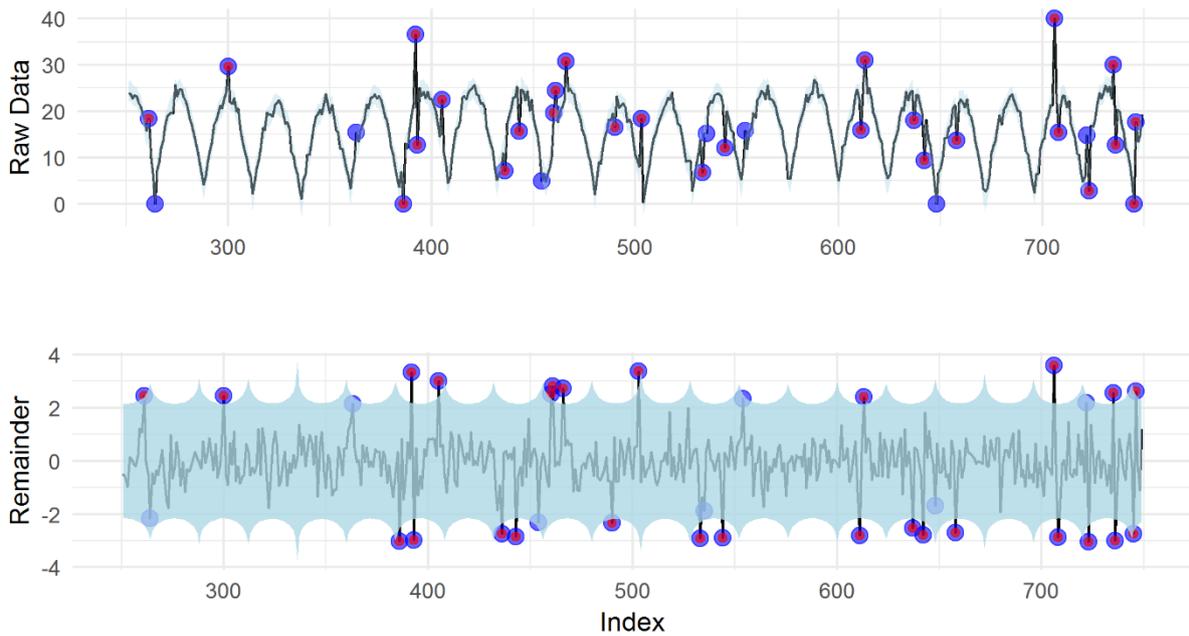

Figure 13. Fluid IQR rule in action (show only a subset of data to illustrate dynamic fence, may need to add)

The dynamic fence associated with our fluid IQR rule helps disregard some outlying observations that would have been identified as relevant outliers by the standard IQR rule. Careful inspection of Figure 13 reveals that most of these observations (depicted only in blue circles) correspond to low activity (fewer sessions). Yet, the fluid IQR rule picks up all other outliers occurring at high activity periods. It shows the ability of the fluid IQR rule to pick more business-relevant outliers.

## 3.2 Comparing outlier detection methods using real data

Figure 14 illustrates the performance of twitter, STL, MSTL and fluid IQR methods in identifying outlying conversion rates in Google merchandise store data. This dataset contains two short periods (mid May and early June) that are significantly different compared to the general conversion pattern. Twitter method is only able to identify these extreme outliers, and fails to identify outliers in the other areas of the time series. This is a serious drawback as occasional extreme events drastically impact the overall performance of the anomaly detection. STL method and then the MSTL method with standard IQR show incremental improvements in detecting outliers in both extreme areas as well as other areas of the time series. Results clearly indicate that the fluid IQR method detects all the extreme outliers, as well as outliers in the other areas of the time series.

Using the Total Absolute Difference in Revenue (TADR) index introduced in section 2.5 and total detected outliers, we assessed how business-relevant each outlier detection method is (Table 5). Results show that the fluid IQR method outperforms the other three methods by a large margin when it comes to business-relevance.

Table 5. Business relevance of outlier detection for four methods

| Method | Total Outliers | TADR ($) |
|---|---|---|
| Twitter | 40 | 25,546 |
| STL | 41 | 26,570 |
| MSTL | 94 | 53,205 |
| Fluid IQR | 162 | 109,925 |

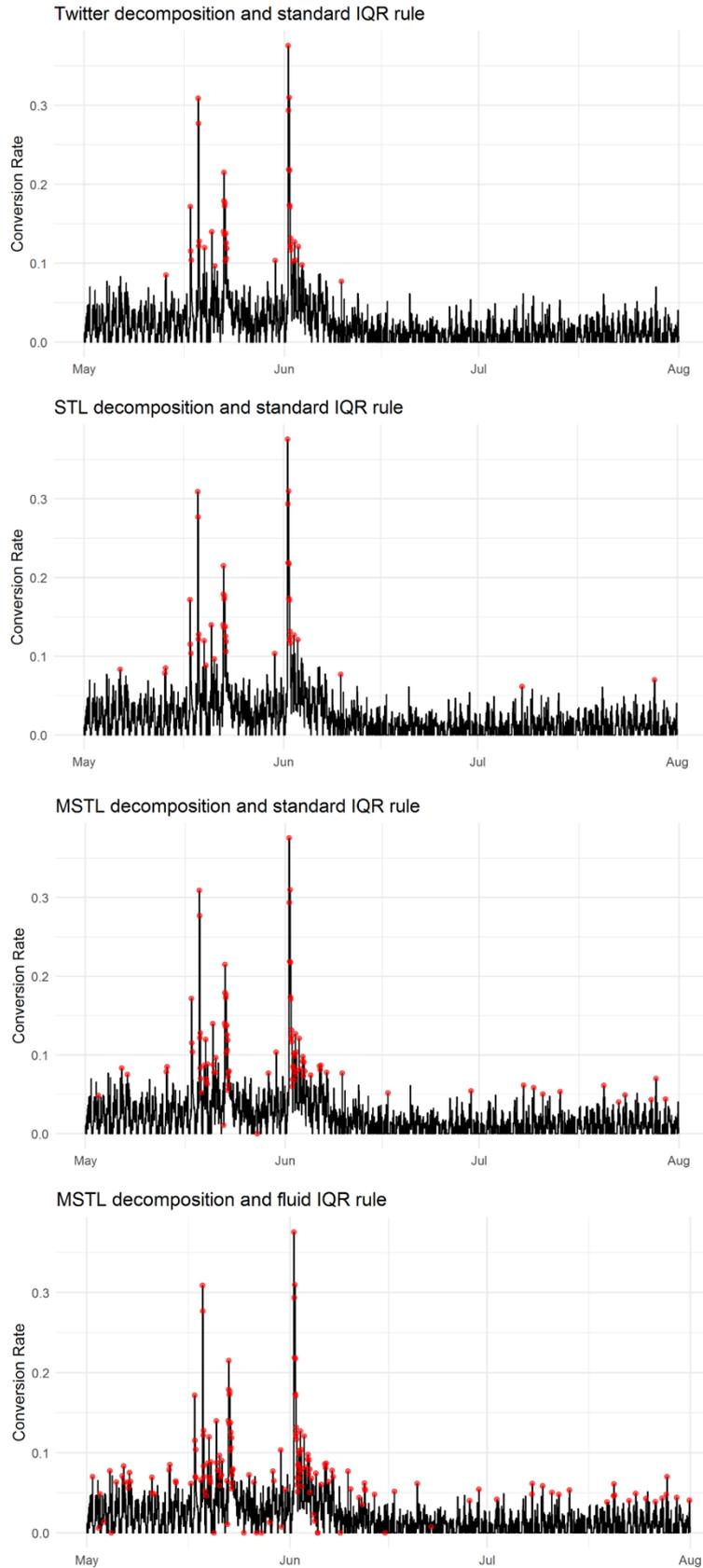

Figure 14. Performance of twitter, STL, MSTL and fluid IQR methods in identifying outlying conversion rates

## 4. Conclusion

In this study, we explored the suitability of various methods to automatically detect outliers in e-commerce conversion rate data. Our approach was to first decompose the conversion time series into trend, seasonality and remainder, and then detecting outliers using the remainder component. We first explained in detail the generation of synthetic conversion data that closely resembles real life conversion rate data. Using the synthetic conversion data, we evaluated three decomposition algorithms: STL with robust LOESS fitting, Twitter method where piece-wise median replaces trend, and MSTL with robust LOESS fitting. We then applied the standard IQR rule (with outer fence) to detect outliers in the remainder component, and demonstrated that MSTL with robust LOESS fitting should be preferred as the decomposition technique for e-commerce conversion data.

An unsupervised outlier detection system that deals with e-commerce conversion data should be intelligently sensitive to the level of activity experienced on the website. For example, an unexpected jump in the conversion rate caused by one extra purchase during typically inactive periods is of low business relevance, while marginally outlying conversions during high activity periods are of significant business value. We presented a novel algorithm called *fluid IQR rule* that achieves the business relevance in an unsupervised anomaly detection algorithm. Fluid IQR method also uses the MSTL decomposition with robust LOESS fitting. However, it then transforms the remainder using inverse hyperbolic sine transformation to minimise any skewness. It further scales the fence values used in IQR based on the level of activity (number of sessions) observed on the e-commerce site. These two enhancements yield an outlying detection method that is far more robust compared to the three methods discussed earlier. When applied on a real e-commerce dataset, fluid IQR method detects outliers that amount to twice the revenue associated with outliers detected by MSTL and standard IQR rule. Future research will evaluate how this new outlier detection method performs in diverse e-commerce business settings.

## Acknowledgements

Authors acknowledge the Innovation Connections grants program of the Australian Federal government for partly funding this research.